\documentclass[10pt,twocolumn]{IEEEtran}

\usepackage{xspace,amsmath,amssymb,amsfonts,epsfig,subfigure,syntonly}
\usepackage{cite,bm,color,url,textcomp,empheq,boxedminipage}
\usepackage{algorithmicx,algorithm}
\usepackage{epstopdf,makecell}
\usepackage{empheq}
\usepackage{pifont}

\usepackage{graphicx,graphics}  
\usepackage{multirow,multicol}
\usepackage{psfrag}    
\usepackage{stfloats}
\usepackage{url}

\newtheorem{lemma}{{Lemma}}[section]

\newtheorem{remark}{{Remark}}[section]

\newcommand{\mv}[1]{\mbox{\boldmath{$ #1 $}}}

\long\def\symbolfootnote[#1]#2{\begingroup
\def\thefootnote{\fnsymbol{footnote}}
\footnote[#1]{#2}\endgroup}

\psfull

\allowdisplaybreaks[4]

\begin{document}

\title{Joint Computation and Communication Cooperation for Energy-Efficient Mobile Edge Computing}

\author{Xiaowen Cao,~\IEEEmembership{Student Member,~IEEE,}~Feng Wang,~\IEEEmembership{Member,~IEEE,}~Jie Xu,~\IEEEmembership{Member,~IEEE,}\\
Rui Zhang,~\IEEEmembership{Fellow,~IEEE,} and Shuguang Cui,~\IEEEmembership{Fellow,~IEEE}

\thanks{Part of this paper has been presented at the International Symposium on Modeling and Optimization in Mobile, Ad Hoc, and Wireless Networks Workshop on Edge and Fog Computing for Intelligent IoT Applications, Shanghai, China, May 7--11, 2018 \cite{Conf_version}.}

\thanks{X. Cao, F. Wang, and J. Xu are with the School of Information Engineering, Guangdong University of Technology, Guangzhou 510006, China (e-mail: caoxwen@outlook.com, fengwang13@gdut.edu.cn, jiexu@gdut.edu.cn).} 

\thanks{R. Zhang is with the Department of Electrical and Computer Engineering, National University of Singapore, Singapore 117583 (e-mail: elezhang@nus.edu.sg).}

\thanks{S. Cui is with the Shenzhen Research Institute of Big Data and School of Science and Engineering, the Chinese University of Hong Kong, Shenzhen 518172, China, and has also been affiliated with Department of Electrical and Computer Engineering, University of California, Davis, CA 95616, USA (e-mail: robert.cui@gmail.com).}

\thanks{Copyright (c) 2012 IEEE. Personal use of this material is permitted. However, permission to use this material for any other purposes must be obtained from the IEEE by sending a request to pubs-permissions@ieee.org.}
}

\maketitle

\begin{abstract}
This paper proposes a novel {\emph{user cooperation}} approach in both computation and communication for mobile edge computing (MEC) systems to improve the energy efficiency for latency-constrained computation. We consider a basic three-node MEC system consisting of a user node, a helper node, and an access point (AP) node attached with an MEC server, in which the user has latency-constrained and computation-intensive tasks to
be executed. We consider two different computation offloading models, namely the partial and binary offloading, respectively. For partial offloading, the tasks at the user are divided into three parts that are executed at the user, helper, and AP, respectively; while for binary offloading, the tasks are executed as a whole only at one of three nodes. Under this setup, we focus on a particular time block and develop an efficient four-slot transmission protocol to enable the \emph{joint computation and communication cooperation}. Besides the local task computing over the whole block, the user can offload some computation tasks to the helper in the first slot, and the helper cooperatively computes these tasks in the remaining time; while in the second and third slots, the helper works as a cooperative relay to help the user offload some other tasks to the AP for remote execution in the fourth slot. For both cases with partial and binary offloading, we jointly optimize the computation and communication resources allocation at both the user and the helper (i.e., the time and transmit power allocations for offloading, and the central process unit (CPU) frequencies for computing), so as to minimize their total energy consumption while satisfying the user's computation latency constraint. Although the two problems are non-convex in general, we develop efficient algorithms to solve them optimally. Numerical results show that the proposed joint computation and communication cooperation approach significantly improves the computation capacity and energy efficiency at the user and helper, as compared to other benchmark schemes without such a joint design.
\end{abstract}


\section{Introduction}\label{sec:introduction}

Recent advancements in the fifth-generation (5G) cellular technologies have enabled various new applications such as the augmented reality (AR), autonomous driving, and Internet of things (IoT). These applications demand ultra-low-latency communication, computation, and control among a large number of wireless devices (e.g., sensors and actuators)\cite{Chiang16}. 
In practice, the real-time computation tasks to be executed can be quite intensive, but wireless devices are generally of small size and only have limited communication, computation, and storage resources (see, e.g., \cite{QianZhang17}). Therefore, how to enhance their computation capabilities and reduce the computation latency is one crucial but challenging issue to be tackled for making these 5G applications a reality.

Conventionally, mobile cloud computing (MCC) has been widely adopted to enhance wireless devices' computation capabilities, by moving their computing and data storage to the remote centralized cloud \cite{Niyato13}. However, as the cloud servers are normally distant from wireless devices, MCC may not be able to meet the stringent computation latency requirements for emerging 5G applications. To overcome such limitations, mobile edge computing (MEC) has been recently proposed as a new solution to provide cloud-like computing at the edge of wireless networks (e.g., access points (APs) and cellular base stations (BSs)), by deploying distributed MEC servers therein \cite{Niyato13,QianZhang17,ETSI,Mach17,Mao17,Bar14,Coz17}. In MEC, wireless devices can offload computation-intensive and latency-critical tasks to APs/BSs in close proximity for remote execution, thus achieving much lower computation latency.

The computation offloading design in MEC systems critically relies on tractable computation task models. Two widely adopted task models in the MEC literature are {\em binary} and {\em partial} offloading, respectively \cite{Mach17, Mao17}. In binary offloading, the computation tasks are not partitionable, and thus should be executed as a whole via either local computing at the user or offloading to the MEC server. This practically corresponds to highly integrated or relatively simple tasks such as those in speech recognition and natural language translation. In contrast, for partial offloading, the computation tasks need to be partitioned into two or more independent parts, which can be executed in parallel by local computing and offloading. This corresponds to applications with multiple fine-grained procedures/components, in, e.g., AR applications \cite{Mao17}. Based on the binary and partial offloading models, the prior works (see, e.g., \cite{Liu16,Niyato15,Bar14,Huang17,Wang17noma,MHchen16CAP,Huang16,Wang17,Suzhi18}) investigated the joint computation and communication resources allocation to improve the performance of MEC. For example, \cite{Liu16} and \cite{Niyato15} considered a single-user MEC system with dynamic task arrivals and channel fading, in which the user jointly optimizes the local computing or offloading decisions to minimize the computation latency, subject to the computation and communication resource constraints. \cite{Huang17,Wang17noma,MHchen16CAP} investigated the energy-efficient design in multiuser MEC systems with multiple users offloading their respective tasks to a single AP/BS for execution, in which the objective is to minimize the users' energy consumption while ensuring their computation latency requirements. Furthermore, \cite{Huang16,Wang17,Suzhi18} proposed wireless powered MEC systems by integrating the emerging wireless power transfer (WPT) technology into MEC for self-sustainable computing, where the AP employs WPT to power the users' local computing and offloading.

Despite the recent research progress, multiuser MEC designs still face several technical challenges. First, the computation resources at the MEC server and the communication resources at the AP should be shared among the actively-computing users. When the user number becomes large, the computation and communication resources allocated to each user are fundamentally limited, thus compromising the benefit of MEC. Next, due to the signal propagation loss over distances, far-apart users may spend much more communication resources than nearby users for offloading, which results in a near-far user fairness issue. Note that the 5G networks are expected to consist of massive wireless devices with certain computation and communication resources. Due to the burst nature of wireless traffic, each active device is highly likely to be surrounded by some idle devices with unused or additional resources. As such, in this paper we propose a novel {\em joint computation and communication cooperation} approach in multiuser MEC systems, such that the nearby users are enabled as helpers to share their computation and communication resources to help actively-computing users, thereby improve the MEC computation performance.

In this paper, we consider a basic three-node MEC system with user cooperation, which consists of a user node, a helper node, and an AP node attached with an MEC server. Here, the helper node can be an IoT sensor, a smart phone, or a laptop, which is nearby and has certain computation and communication resources.\footnote{In general, the computation capability of the helper should be comparable or stronger than that of the user in order for the computation cooperation to be feasible.} We focus on the user's latency-constrained computation within a given time block. To implement the joint computation and communication cooperation, the block is divided into four time slots. Specifically, in the first slot, the user offloads some computation tasks to the helper for remote execution. In the second and third slots, the helper acts as a decode-and-forward (DF) relay to help the user offload some other computation tasks to the AP, for remote execution at the MEC server in the fourth slot. Under this setup, we pursue an energy-efficient user cooperation MEC design for both the partial and binary offloading cases, by jointly optimizing the computation and communication resource allocations. The main results of this paper are summarized as follows.
\begin{itemize}

\item First, for the partial offloading case, the user's computation tasks are partitioned into three parts for local computation, offloading to helper, and offloading to AP, respectively. Towards minimizing the total energy consumption at both the user and the helper subject to the user's computation latency constraint, we jointly optimize the task partition of the user, the central process unit (CPU) frequencies for local computing at both the user and the helper, as well as the time and transmit power allocations for offloading. The non-convex problem of interests in general can be reformulated into a convex one. Leveraging the Lagrange duality method, we obtain the globally optimal solution in a semi-closed form.

\item Next, for the binary offloading case, the user should execute the non-partitionable computation tasks by choosing one among three computation modes, i.e., the local computing at the user, computation cooperation (offloading to the helper), and communication cooperation (offloading to the AP). Solving the resultant latency-constrained energy minimization problem, we develop an efficient optimal algorithm, by firstly choosing the computation mode and then optimizing the corresponding joint computation and communication resources allocation.

\item Finally, extensive numerical results show that the proposed joint computation and communication cooperation approach achieves significant performance gains in terms of both the computation capacity and the system energy efficiency, compared with other benchmark schemes without such a joint design.
\end{itemize}

It is worth noting that there have been prior studies on communication cooperation (see, e.g., \cite{Sen03,Nosratinia04,Tse04,Veeravalli05,Rui14,kkwong18,Li12}) or computation cooperation \cite{MingjunXiao15,ZSheng18,Mti13,xuchenfog,kaibinhuang17}, respectively. On one hand, the cooperative communication via relaying has been extensively investigated in wireless communication systems to increase the communication rate and improve the communication reliability \cite{Tse04,Veeravalli05}, and applied in various other setups such as the wireless powered communication \cite{Rui14} and the wireless powered MEC systems \cite{kkwong18}. On the other hand, cooperative computation has emerged as a viable technique in MEC systems, which enables end users to exploit computation resources at nearby wireless devices (instead of APs or BSs). For example, in the so-called device-to-device (D2D) fogging \cite{xuchenfog} and peer-to-peer (P2P) cooperative computing \cite{kaibinhuang17}, users with intensive computing tasks can offload all or part of the tasks to other idle users via D2D or P2P communications for execution. Similar computation task sharing among wireless devices has also been investigated in mobile social networks with crowdsensing \cite{MingjunXiao15} and in mobile wireless sensor networks \cite{ZSheng18,Mti13} for data fusion. However, different from these existing works with either communication or computation cooperation, this work is the first to pursue a {\em joint computation and communication cooperation} approach, by unifying both of them for further improving the user's computation performance. Also note that this work with user cooperation is different from the prior works on multiuser MEC systems \cite{Huang17,Wang17noma,MHchen16CAP}, in which multiple users offload their own computation tasks to the AP/BS for execution, without user cooperation considered.

The remainder of this paper is organized as follows. Section~\ref{sec:system1} introduces the system model. Section~\ref{sec:formulation} formulates the latency-constrained energy minimization problems under the partial and binary offloading models, respectively. Sections~\ref{sec:partial} and~\ref{sec:binary} present the optimal solutions to the two problems of our interests, respectively. Section~\ref{sec:simulation} provides numerical results, followed by the conclusion in Section~\ref{sec:conclusions}.

\section{System Model}\label{sec:system1}

\begin{figure}
\centering
  \includegraphics[width=8.0cm]{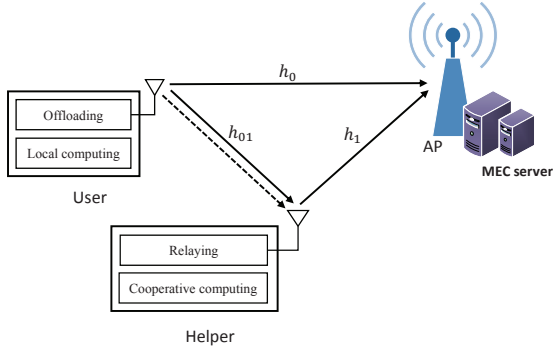}
\caption{A basic three-node MEC system with joint computation and communication cooperation. The dashed and solid lines indicate the tasks offloaded to the helper (for computation cooperation) and to the AP (via the helper's communication cooperation as a relay), respectively.} \label{fig:1}
\end{figure}

As shown in Fig.~\ref{fig:1}, we consider a basic three-node MEC system consisting of one user node, one helper node, and one AP node with an MEC server integrated, in which the three nodes are each equipped with one single antenna. We focus on a time block with duration $T>0$, where the user needs to successfully execute computation tasks with $L>0$ task input-bits within this block. By considering a latency-critical application, we assume that the block duration $T$ is smaller than the channel coherence time, such that the channel power gain remains unchanged within the block of interest. Such an assumption has been commonly adopted in prior works \cite{Huang17,Wang17noma,MHchen16CAP,Huang16,Suzhi18,Wang17}. It is further assumed that there is a central controller that is able to collect the global channel state information (CSI), and computation-related information for the three nodes; accordingly, the central controller can design and coordinate the computation and communication cooperation among the three nodes. This serves as a performance upper bound (or energy consumption lower bound) for practical cases when only partial CSI and computation-related information are known.


Specifically, without loss of generality, the $L$ task input-bits can be divided into three parts intended for local computing, offloading to helper, and offloading to AP, respectively. Let $l_u \ge 0$, $l_h \ge 0$, and $l_a \ge 0$ denote the numbers of task input-bits for local computing at the user, offloading to the helper, and offloading to the AP, respectively. We then have
\begin{equation}\label{eqn:L}
l_u+l_h+l_a = L.
\end{equation}
Consider the two cases with partial offloading and binary offloading, respectively. In partial offloading, the computation task can be arbitrarily partitioned into subtasks. By assuming the number of subtasks are sufficiently large in this case, it is reasonable to approximate $l_u$, $l_h$, and $l_a$ as real numbers between 0 and $L$ subject to \eqref{eqn:L}. In binary offloading, $l_u$, $l_h$, and $l_a$ can only be set as 0 or $L$, and there is only one variable among them equal to $L$ due to \eqref{eqn:L}.
\begin{figure}
\centering
 \includegraphics[width=8cm]{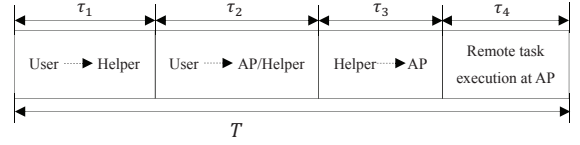}
\caption{MEC protocol with joint computation and communication cooperation.} \label{fig:2}
\end{figure}

\subsection{MEC Protocol With Joint Computation and Communication Cooperation}

As shown in Fig.~\ref{fig:2}, the duration-$T$ block is generally divided into four slots for joint computation and communication cooperation. In the first slot with duration $\tau_1\ge 0$, the user offloads the $l_h$ task input-bits to the helper, and the helper can then execute them in the remaining time with duration $T - \tau_1$. In the second and third slots, the helper acts as a DF relay to help the user offload $l_a$ task input-bits to the AP. In the second slot with duration $\tau_2\ge 0$, the user transmits wireless signals containing the $l_a$ task input-bits to both the AP and the helper simultaneously. After successfully decoding the received task input-bits, the helper forwards them to the AP in the third slot with duration $\tau_3\ge 0$. After decoding the signals from the user and the helper, the MEC server can remotely execute the offloaded tasks in the fourth time slot with duration $\tau_4\ge 0$.

As the computation results are normally of much smaller size than the input bits, the time for downloading the results to the user is negligible compared to the offloading time. Thus, we ignore the downloading time in this paper. In order to ensure the computation tasks to be successfully executed before the end of this block, we have the following time constraint
\begin{equation}\label{eqn:T}
\tau_1+\tau_{2}+\tau_{3}+\tau_4 \leq T.
\end{equation}

\subsection{Computation Offloading}
In this subsection, we discuss the computation offloading from the user to the helper and the AP, respectively.
\subsubsection{Computation Offloading to Helper}
In the first slot, the user offloads $l_h$ task input-bits to the helper with transmit power $P_1\ge 0$. Let $h_{01}>0$ denote the channel power gain from the user to the helper, and $B$ the system bandwidth. Accordingly, the achievable data rate (in bits/sec) for offloading from the user to the helper is given by
\begin{align}\label{eqn:r01}
r_{01}(P_1)=B\log_2\left(1+\frac{P_1h_{01}}{\Gamma\sigma_1^2}\right),
\end{align}
where $\sigma_1^2$ represents the power of additive white Gaussian noise (AWGN) at the helper, and $\Gamma\ge1$ is a constant term accounting for the gap from the channel capacity due to a practical modulation and coding scheme (MCS). For simplicity, $\Gamma=1$ is assumed throughout this paper. Consequently, we have the number $l_h$ of task input-bits as
\begin{align}\label{eq:l1}
l_h = \tau_1r_{01}(P_1).
\end{align}
Furthermore, let $P_{u,\max}$ denote the maximum transmit power at the user, and thus we have $0\leq P_1 \le P_{u,\max}$. For computation offloading, we consider the user's transmission energy as the dominant energy consumption and ignore the energy consumed by circuits in its radio-frequency (RF) chains, baseband signal processing, etc. Therefore, in the first slot, the energy consumption for the user to offload $l_h$ task input-bits to the helper is given by
\begin{align}\label{eqn:par:E1off}
E_1^{\rm offl}=\tau_1P_1.
\end{align}

\subsubsection{Computation Offloading to AP Assisted by Helper}

In the second and third slots, the helper acts as a DF relay to help the user offload $l_a$ task input-bits to the AP. Denote by $0\leq P_2\le P_{u,\max}$ the user's transmit power in the second slot. In this case, the achievable data rate from the user to the helper is given by $r_{01}(P_2)$ with $r_{01}(\cdot)$ defined in (\ref{eqn:r01}). Denoting $h_0>0$ as the channel power gain from the user to the AP, the achievable data rate from the user to the AP is
\begin{equation}
r_0(P_2) = B\log_2\left(1+\frac{P_2h_{0}}{\sigma_0^2}\right),
\end{equation}
where $\sigma_0^2$ is the noise power at the AP receiver.

After successfully decoding the received message, the helper forwards it to the AP in the third slot with the transmit power $0\leq P_3 \leq P_{h,\max}$, where $P_{h,\max}$ denotes the maximum transmit power at the helper. Let $h_{1}>0$ denote the channel power gain from the helper to the AP. The achievable data rate from the helper to the AP is thus
\begin{equation}
r_1(P_3)=B\log_2\left(1+\frac{P_3h_{1}}{\sigma_0^2}\right).
\end{equation}

By combining the second and third slots, the number of $l_a$ task input-bits offloaded to the AP via a DF relay (the helper) should satisfy \cite{Veeravalli05,Tse04,Sen03}
\begin{align}\label{eq:l2}
l_a = \min\left(\tau_2r_0(P_2)+\tau_3r_1(P_3),~\tau_2r_{01}(P_2)\right).
\end{align}
As in (\ref{eqn:par:E1off}), we consider the user's and helper's transmission energy consumption for offloading as the dominant energy consumption in both the second and third slots. Therefore, we have
\begin{align}
E_2^{\rm offl} &=\tau_2P_2\label{eqn:par:E2off}\\
E_3^{\rm offl} &=\tau_3P_3.\label{eqn:par:E3off}
\end{align}

\subsection{Computing at User, Helper, and AP}
In the subsection, we explain the computing models at the user, the helper, and the AP, respectively.

\subsubsection{Local Computing at User}
The user executes the computation tasks with $l_u$ task input-bits within the whole block. In practice, the number of CPU cycles for executing a computation task is highly dependent on various factors such as the specific applications, the number of task input-bits, as well as the hardware (e.g., CPU and memory) architectures at the computing device \cite{kappa}. To characterize the most essential computation and communication tradeoff and as commonly adopted in the literature (e.g., \cite{Liu16,Huang16,Huang17,Niyato15,Wang17,Wang17noma,MHchen16CAP}), we consider that the number of CPU cycles for this task is a linear function with respect to the number of task input-bits, where $c_u$ denotes the number of CPU cycles for computing each one task input-bit at the user. Also, let $f_{u,n}$ denote the CPU frequency for the $n$-th cycle, where $n\in\{1,\ldots,c_u l_u\}$. Note that the CPU frequency $f_{u,n}$ is constrained by a maximum value, denoted by $f_{u,\max}$, i.e.,
\begin{equation}\label{eqn:f_u}
f_{u,n}\leq f_{u,\max},~~\forall n\in\{1,\ldots,c_u l_u\}.
\end{equation}
As the local computing for the $l_u$ task input-bits should be successfully accomplished before the end of the block, we have the following computation latency requirement
\begin{equation}\label{eqn:latency}
\sum_{n=1}^{c_ul_u}\frac{1}{f_{u,n}}\leq T.
 \end{equation}
Accordingly, the user's energy consumption for local computing is \cite{Mao17,Wang17}
\begin{equation}
E_u^{\rm comp} = \sum_{n=1}^{c_ul_u}\kappa_uf_{u,n}^2,\label{eqn:energy:local:user}
\end{equation}
where $\kappa_u$ denotes the effective capacitance coefficient that depends on the chip architecture at the user \cite{kappa}. It has been shown in \cite[Lemma~1]{Wang17} that to save the computation energy consumption with a computation latency requirement, it is optimal to set the CPU frequencies to be identical for different CPU cycles. By using this fact and letting the constraint in (\ref{eqn:latency}) be met with strict equality (for minimizing the computation energy consumption), we have
\begin{equation}\label{eqn:f_u_max}
f_{u,1}=f_{ u,2}=...=f_{ u,c_ul_u}=c_ul_u/T.
\end{equation}
Substituting \eqref{eqn:f_u_max} into (\ref{eqn:energy:local:user}), the user's corresponding energy consumption for local computing is re-expressed as
\begin{align}\label{eqn:par:Eu_comp}
E_u^{\rm comp}=\frac {\kappa_uc_u^3l_u^3}{T^2}.
\end{align}
Combining \eqref{eqn:f_u_max} with the maximum CPU frequency constraint in \eqref{eqn:f_u}, it follows that
\begin{align}\label{eq:f0:max}
c_ul_u\leq T f_{u,\max}.
\end{align}

\subsubsection{Cooperative Computing at Helper}

After receiving the offloaded $l_h$ task input-bits in the first time slot, the helper executes the tasks during the remaining time with duration $(T-\tau_1)$. Let $f_{h,n}$ and $f_{h,\max}$ denote the CPU frequency for the $n$-th CPU cycle and the maximum CPU frequency at the helper, respectively. Similarly as for the local computing at the user, it is optimal for helper to set the CPU frequency for the $n$-th CPU cycle as $f_{h,n}=c_hl_h/(T-\tau_1)$, $n\in\{1,\ldots,c_hl_h\}$, where $c_h$ is the number of CPU cycles for computing one task-input bit at the helper. Accordingly, the energy consumption for cooperative computation at the helper is given by
\begin{equation}\label{eqn:par:Eh_comp}
E_h^{\rm comp}=\frac {\kappa_hc_h^3l_h^3}{(T-\tau_1)^2},
\end{equation}
where $\kappa_h$ is the effective capacitance coefficient of the helper.

Similarly as in \eqref{eq:f0:max}, we have the constraint on the number of task input-bits as
\begin{align}\label{eq:f1:max}
c_hl_h\leq (T-\tau_1)f_{h,\max},
\end{align}
where $f_{h,\max}$ denotes the maximum CPU frequency for the helper.

\subsubsection{Remote Computing at AP (MEC Server)}
In the fourth slot, the MEC server at the AP executes the offloaded $l_a$ task input-bits. In order to minimize the remote execution, the MEC server executes the offloaded tasks at its maximal CPU frequency, denoted by $f_{a,\max}$. Hence, the time duration $\tau_4$ for the MEC server to execute the $l_a$ offloaded bits is
\begin{align}\label{eq:fa:max}
\tau_4 =c_al_a/f_{a,\max},
\end{align}
where $c_a$ represents the number of CPU cycles required for computing one task-input bit at the AP.
By substituting \eqref{eq:fa:max} into \eqref{eqn:T}, the time allocation constraint is re-expressed as
\begin{align}\label{eqn:T1}
\tau_1 + \tau_2 + \tau_3 + c_al_a/f_{a,\max} \le T.
\end{align}

\section{Problem Formulation}\label{sec:formulation}
In this paper, we pursue an energy-efficient design for the three-node MEC system. As the AP normally has reliable power supply, we focus on the energy consumption at the wireless devices side (i.e., the user and helper) as the performance metric. In particular, we aim to minimize the total energy consumption at both the user and the helper (i.e., $\sum_{i=1}^3 E_i^{\rm offl} + E_u^{\rm comp} + E_h^{\rm comp}$), subject to the user's computation latency constraint $T$, by optimizing the task partition of the user, as well as the joint computation and communication resources allocation. The design variables include the time allocation vector of the slots ${\bm \tau}\triangleq [\tau_1,\tau_2,\tau_3]$, the user's task partition vector ${\bm l}\triangleq [l_u,l_h,l_a]$, and the transmit power allocation vector ${\bm P}\triangleq [P_1,P_2,P_3]$ for offloading of the user and helper.

In the case with partial offloading, the latency-constrained energy minimization problem is formulated as
\begin{subequations}\label{eq.part_prob}
\begin{align}
({\rm P1}):~\min_{{\bm P}, {\bm \tau}, {\bm l}} &~\frac {\kappa_uc_u^3l_u^3}{T^2}+\frac {\kappa_hc_h^3l_h^3}{(T-\tau_1)^2}+\sum_{i=1}^3\tau_iP_i \label{eq.part_prob.a}\\
{\rm s.t.} &~l_h \leq \tau_1r_{01}(P_1)\label{eqn:l1}\\
&~ l_a \leq \min\big(\tau_2r_0(P_2)+\tau_3r_1(P_3),~\tau_2r_{01}(P_2)\big)\label{eqn:l2}\\
&~0\leq P_j\leq P_{u,\max},~\forall j\in \{1, 2\} \label{eq.part_prob.b}\\
&~0\leq P_3\leq P_{h,\max} \label{eq.part_prob.b1}\\
& ~0\leq \tau_i\leq T, ~\forall i\in\{1,2,3\} \label{eq.part_prob.c}\\
&~ l_u \ge 0,~ l_h \ge 0,~ l_a \ge 0  \label{eq.part_prob.d}\\
&~\eqref{eqn:L},~\eqref{eq:f0:max},~\eqref{eq:f1:max},~{\text{and}}~\eqref{eqn:T1}, \notag
\end{align}
\end{subequations}
where (\ref{eqn:L}) denotes the task partition constraint, \eqref{eq:f0:max} and \eqref{eq:f1:max} are the maximum CPU frequency constraints at the user and the helper, respectively, (\ref{eqn:T1}) denotes the time allocation constraint, (\ref{eqn:l1}) and (\ref{eqn:l2}) denote the constraints for the numbers of the offloaded bits from the user to the helper and to the AP, respectively. Note that in problem (P1), we replace the two equalities in (\ref{eq:l1}) and (\ref{eq:l2}) as two inequality constraints (\ref{eqn:l1}) and (\ref{eqn:l2}). It is immediate that constraints (\ref{eqn:l1}) and (\ref{eqn:l2}) should be met with strict equality at optimality of problem (P1). Also note that problem (P1) is non-convex, due to the coupling of $\tau_i$ and $P_i$ in the objective function \eqref{eq.part_prob.a} and the constraints \eqref{eqn:l1} and \eqref{eqn:l2}. Nonetheless, in Section~\ref{sec:partial} we will transform (P1) into an equivalent convex problem and then present an efficiently algorithm to obtain the optimal solution of problem (P1) in a semi-closed form.

In the case with binary offloading, the latency-constrained energy minimization problem is formulated as
\begin{subequations}\label{eq.part_prob2}
\begin{align}
({\rm P2}):~\min_{{\bm P}, {\bm \tau}, {\bm l}} &~\frac {\kappa_uc_u^3l_u^3}{T^2}+\frac {\kappa_hc_h^3l_h^3}{(T-\tau_1)^2}+\sum_{i=1}^3\tau_iP_i\label{eq.part_prob2.a}\\
{\rm s.t.} &~l_u\in\{0,L\},~ l_h\in\{0,L\},~l_a\in\{0,L\}\label{eq.part_prob2.b}\\
&~\eqref{eqn:L},~\eqref{eq:f0:max},~\eqref{eq:f1:max},~\eqref{eqn:T1},~{\text{and}}~(\ref{eq.part_prob}\text{b--f}).\notag
\end{align}
\end{subequations}
Note that problem (P2) is a mixed-integer nonlinear program (MINLP) \cite{MINLP} due to the involvement of integer variables $l_u$, $l_h$, and $l_a$. In Section \ref{sec:binary}, we will develop an efficient algorithm to solve problem (P2) optimally by examining three computation modes, respectively.

\subsection{Feasibility of $\rm (P1)$ and $\rm (P2)$}

Before solving problems (P1) and (P2), we first check their feasibility to determine whether the MEC system of interests can support the latency-constrained task execution or not. Let $L^{(1)}_{\max}$ and $L^{(2)}_{\max}$ denote the maximum numbers of task input-bits supported by the MEC system within the duration-$T$ block under the partial and binary offloading cases, respectively. Evidently, if $L_{\max}^{(1)} \ge L$ (or $L_{\max}^{(2)} \ge L$), then problem (P1) (or (P2)) is feasible; otherwise, the corresponding problem is not feasible. Therefore, the feasibility checking procedures of problems (P1) and (P2) correspond to determining $L^{(1)}_{\max}$ and $L^{(2)}_{\max}$, respectively.

First, consider the partial offloading case. The maximum number $L^{(1)}_{\max}$ of task input-bits is attained when the three nodes fully use their available communication and computation resources. This corresponds to setting as $P_1=P_2= P_{u,\max}$, $P_3= P_{h,\max}$, and letting the constraints \eqref{eq:f0:max}, \eqref{eq:f1:max}, \eqref{eqn:T1}, and \eqref{eqn:l2} be met with the strict equality in problem (P1). As a result, $L_{\max}^{(1)}$ is the optimal value of the following problem:
\begin{align}\label{eq.part_feaprob1}
L_{\max}^{(1)} \triangleq \max_{{\bm \tau}, {\bm l}} &~l_u+l_h+l_a \\
{\rm s.t.} &~ \tau_1 + \tau_2 + \tau_3 + c_al_a/f_{a,\max} =T \notag \\
&~l_h\leq \tau_1r_{01}(P_{u,\max}),~l_a\leq \tau_2r_{01}(P_{u,\max}) \nonumber\\
&~ c_u l_u/T = f_{u,\max},~ c_hl_h/(T-\tau_1)= f_{h,\max} \nonumber\\
&~\tau_2r_0(P_{u,\max})+\tau_3r_1(P_{h,\max})=\tau_2r_{01}(P_{u,\max})\nonumber\\
&~\eqref{eq.part_prob.c}~{\text{and}}~\eqref{eq.part_prob.d}.\nonumber
\end{align}
Note that problem (\ref{eq.part_feaprob1}) is a linear program (LP) and can thus be efficiently solved via standard convex optimization techniques such as the interior point method \cite{Boyd2004}. By comparing $L_{\max}^{(1)}$ versus $L$, the feasibility of problem (P1) is checked.

Next, consider the binary offloading case. The user's computation tasks can only be executed by one of the three computation modes, namely the {\em local computing}, {\em computation cooperation} (offloading to helper), and {\em communication cooperation} (offloading to AP). For the three modes, the maximum numbers of supportable task input-bits can be readily obtained, as stated in the following.
\begin{itemize}
  \item For the local-computing mode, we have $l_h = l_a = 0$. With the maximum CPU frequency $f_{u,\max}$ and setting \eqref{eq:f0:max} to be tight, the maximum supportable number of task input-bits is given by
      \begin{align}\label{eq.bin_lumax}
      l_{u,\max}^{(2)}=\frac{Tf_{u,\rm max}}{c_u}.
      \end{align}
  \item For the computation-cooperation mode, we have $l_u = l_a = 0$. By setting the user's transmit power as $P_1 = P_{u,\max}$, and making the constraints \eqref{eq:f1:max} and \eqref{eqn:l1} be tight in problem (P2), the maximum number of task input-bits is thus
      \begin{align}\label{eq.bin_lhmax}
      l_{h,\max}^{(2)} = \tau_1^br_{01}(P_{u,\max}),
      \end{align}
      where $\tau_1^b={Tf_{h,\max}}/({c_hr_{01}(P_{u,\max})+f_{h,\max}})$ is the user's optimal time allocation for offloading.
  \item For the communication-cooperation mode, we have $l_u = l_h = 0$, $P_2=P_{u,\max}$, and $P_3=P_{h,\max}$ in problem (P2). The maximum number of task input-bits $l^{(2)}_{a,\max}$ is obtained by solving the following LP:
        \begin{align}\label{eq.bin_lamax}
l^{(2)}_{a,\max}= \max_{\tau_2,\tau_3,l_a} &~l_a \\
        {\rm s.t.} & ~l_a\leq \min \Big(\tau_2r_{01}(P_{u,\max}),\notag\\
        &\quad\quad\quad \tau_2r_0(P_{u,\max})+\tau_3r_1(P_{h,\max})\Big)\nonumber\\
        &~ \tau_2+\tau_3+c_al_a/f_{a,\max} \leq T\nonumber\\
        &~0\le\tau_2\le T, ~0\le\tau_3 \le T.\nonumber
        \end{align}
\end{itemize}

Based on \eqref{eq.bin_lumax}--\eqref{eq.bin_lamax}, the maximum number of supportable task input-bits for the binary offloading case is given by
\begin{align}
L_{\max}^{(2)} =\max\left( l_{u,\max}^{(2)},l_{h,\max}^{(2)},l_{a,\max}^{(2)} \right).
\end{align}
By comparing $L_{\max}^{(2)}$ with $L$, the feasibility of problem (P2) is checked.

By comparing $L_{\max}^{(1)}$ and $L_{\max}^{(2)}$, we show that $L_{\max}^{(1)} \ge L_{\max}^{(2)}$. This is expected since that any feasible solution to problem (P2) is always feasible for problem (P1), but the reverse is generally not true. In other words, the partial offloading case can better utilize the distributed computation resources at different nodes, and thus achieves higher computation capacity than the binary offloading case. 


\section{Optimal Solution to ({\rm P1})}\label{sec:partial}

In this section, we present an efficient algorithm for optimally solving problem (P1) in the partial offloading case.

 Towards this end, we introduce an auxiliary variable vector ${\bm E}\triangleq [E_1,E_2,E_3]$ with $E_i = P_i\tau_i$ for all $i\in\{1,2,3\}$. Then it holds that $P_i=E_i/\tau_i$ if $\tau_i > 0$, and $P_i=0$ if either $E_i=0$ or $\tau_i=0$ for any $i\in\{1,2,3\}$. By substituting $P_i=E_i/\tau_i$, $i\in\{1,2,3\}$, problem (P1) can be reformulated as
\begin{subequations}\label{eq.part_prob.1}
\begin{align}
({\rm P1.1}):~\min_{{\bm E}, {\bm \tau}, {\bm l}} &~~\frac {\kappa_uc_u^3l_u^3}{T^2}+\frac {\kappa_hc_h^3l_h^3}{(T-\tau_1)^2}+\sum_{i=1}^3E_i \label{eq.part_prob.1.a}\\
{\rm s.t.} &~l_h\leq \tau_1r_{01}\left(\frac{E_1}{\tau_1}\right)\label{eq.part_prob.1.b} \\
&~ l_a \leq \tau_2r_0\left(\frac{E_2}{\tau_2}\right)+\tau_3r_1\left(\frac{E_3}{\tau_3}\right)\label{eq.part_prob.1.c}\\
&~l_a \leq  \tau_2r_{01}\left(\frac{E_2}{\tau_2}\right)\label{eq.part_prob.1.d}\\
&~ 0\leq E_j\leq \tau_jP_{u,\max},~~\forall j\in\{1,2\} \label{eq.part_prob.1.e}\\
&~ 0\leq E_3\leq \tau_3P_{h,\max} \label{eq.part_prob.1.e1}\\
&~\eqref{eqn:L},~\eqref{eq:f0:max},~\eqref{eq:f1:max},~\eqref{eqn:T1},~\eqref{eq.part_prob.c},~{\text{and}}~\eqref{eq.part_prob.d}\nonumber,
\end{align}
\end{subequations}
where both \eqref{eq.part_prob.1.c} and \eqref{eq.part_prob.1.d} follow from \eqref{eqn:l2}.
\begin{lemma}\label{lem.convex}
Problem (P1.1) is a convex problem.
\end{lemma}

\begin{IEEEproof}
The function $r_j(x)$ is a concave function with respect to $x \ge 0$ for any $j\in\{0,1,01\}$, and thus its perspective function $x r_j\left(\frac{y}{x}\right)$ is jointly concave with respect to $x> 0$ and $y\ge 0$ \cite{Boyd2004}. As a result, the set defined by constraints (\ref{eq.part_prob.1.b})--(\ref{eq.part_prob.1}d) becomes convex. The function $l^3/\tau^2$ is a convex function with respect to $l \ge 0$ and $\tau > 0$, and hence the term ${\kappa_hc_h^3l_h^3}/{(T-\tau_1)^2}$ in the objective function is jointly convex with respect to $l_h\ge 0$ and $0\le \tau_1<T$. Therefore, problem (P1.1) is convex.
\end{IEEEproof}

As stated in Lemma~\ref{lem.convex}, problem (P1.1) is convex and can thus be optimally solved by the standard interior point method \cite{Boyd2004}. Alternatively, to gain essential engineering insights, we next leverage the Lagrange duality method to obtain a well-structured optimal solution for problem (P1.1).

Let $\lambda_1\ge 0$, $\lambda_2\ge 0$, and $\lambda_3\ge 0$ denote the dual variables associated with the constraints in (\ref{eq.part_prob.1}\text{b--d}), respectively, $\mu_1 \ge 0$ and $ \mu_2\in\mathbb{R}$ be the dual variables associated with the constraints in \eqref{eqn:T1} and \eqref{eqn:L}, respectively. Define ${\bm \lambda}\triangleq [\lambda_1, \lambda_2, \lambda_3]$ and ${\bm \mu}\triangleq [\mu_1, \mu_2]$. The partial Lagrangian of problem (P1.1) is given by
\begin{align*}
&{{\cal L}}({\bm E},{\bm \tau}, {\bm l}, {\bm \lambda}, {\bm \mu})
= \sum_{i=1}^3E_i+\mu_1\tau_1+\frac {\kappa_hc_h^3l_h^3}{(T-\tau_1)^2}+(\lambda_1-\mu_2)l_h \\
&~~ -\lambda_1\tau_1r_{01}\left(\frac{E_1}{\tau_1}\right)-\lambda_2\tau_2r_0\left(\frac{E_2}{\tau_2}\right)-\lambda_3\tau_2r_{01}\left(\frac{E_2}{\tau_2}\right)\\
&~~+\mu_1\tau_2-\lambda_2\tau_3r_1\left(\frac{E_3}{\tau_3} \right)+\mu_1\tau_3+\frac {\kappa_uc_u^3l_u^3}{T^2}-\mu_2l_u\\
 &~~+\left(\lambda_2+\lambda_3+\mu_1c_a/f_{a,\max}-\mu_2\right)l_a-\mu_1T+\mu_2 L.
\end{align*}
Then the dual function of problem (P1.1) is given by
\begin{align}\label{eq.dual_function}
g( {\bm \lambda}, {\bm \mu}) = \min_{{\bm E},{\bm \tau}, {\bm l}} &~{{\cal L}}({\bm E},{\bm \tau}, {\bm l}, {\bm \lambda}, {\bm \mu})\\
{\rm s.t.}&~\eqref{eq:f0:max},~\eqref{eq:f1:max},~\eqref{eq.part_prob.c},~\eqref{eq.part_prob.d},~\eqref{eq.part_prob.1.e},~{\text{and}}~\eqref{eq.part_prob.1.e1}. \nonumber
 \end{align}

\begin{lemma}\label{lem0}
In order for the dual function $g(\mv \lambda, \mv \mu)$ to be bounded from below, it must hold that $\lambda_1-\mu_2\ge0~{\text {and}} ~\lambda_2+\lambda_3+\mu_1c_a/f_{a,\max}-\mu_2 \ge 0$.
\end{lemma}
\begin{IEEEproof}
See Appendix~\ref{AppendixA}.
\end{IEEEproof}

Based on Lemma \ref{lem0}, the dual problem of problem (P1.1) is given by
\begin{subequations}\label{eq.dual_prob}
\begin{align}
{\rm (D1.1)}:~\max_{{\bm \lambda}, {\bm \mu}} &\quad g({\bm \lambda},{\bm \mu})\\
{\rm s.t.}&~~ \mu_1 \geq 0,~\lambda_i\ge 0,~~\forall i\in\{1,2,3\} \\
&~~\lambda_1-\mu_2\ge0\\
&~~\lambda_2+\lambda_3+\mu_1c_a/f_{a,\max}-\mu_2 \ge 0.
\end{align}
\end{subequations}
Denote $\cal X$ and $(\bm \lambda^{\rm opt1},\bm \mu^{\rm opt1})$ as the feasible set and the optimal solution of $({\bm \lambda},{\bm \mu})$ for problem (D1.1), respectively.

Since problem $({\rm P1.1})$ is convex and satisfies the Slater's condition, strong duality holds between problems $({\rm P1.1})$ and $({\rm D1.1})$\cite{Boyd2004}. As a result, one can solve problem $({\rm P1.1})$ by equivalently solving its dual problem $({\rm D1.1})$. In the following, we first evaluate the dual function $g({\bm \lambda},{\bm \mu})$ under any given $({\bm \lambda},{\bm \mu})\in \cal X$, and then obtain the optimal dual variables $(\bm \lambda^{\rm opt1},\bm \mu^{\rm opt1})$ to maximize $g({\bm \lambda},{\bm \mu})$. Denote $({\bm E}^*,{\bm \tau}^*,{\bm l}^*)$ as the optimal solution to problem (\ref{eq.dual_function}) under any given $({\bm \lambda},{\bm \mu})\in \cal X$, $({\bm E}^{\rm opt1},{\bm \tau}^{\rm opt1},{\bm l}^{\rm opt1})$ as the optimal primal solution to problem $({\rm P1.1})$.


\subsection{Derivation of Dual Function $g(\mv \lambda, \mv \mu)$}\label{sec:evalution}

First, we obtain $g(\mv \lambda, \mv \mu)$ by solving (\ref{eq.dual_function}) under any given $(\mv \lambda,\mv \mu)\in{\cal X}$. Equivalently, (\ref{eq.dual_function}) can be decomposed into the following five subproblems.
\begin{align}
\min_{E_1, \tau_1, l_h} & E_1+\mu_1\tau_1-\lambda_1\tau_1r_{01}\left(\frac{E_1}{\tau_1}\right)+\frac{\kappa_hc_h^3l_h^3}{(T-\tau_1)^2} +(\lambda_1-\mu_2)l_h \notag \\
~~{\rm s.t.}&~\eqref{eq:f1:max},~0\leq E_1\leq \tau_1P_{u,\max},~0\leq \tau_1\leq T,~l_h \ge 0. \label{eq.sub_part_prob1}
\end{align}
\begin{align}
&\min_{E_2, \tau_2 } ~ E_2+\mu_1\tau_2-\lambda_2\tau_2r_0\left(\frac{E_2}{\tau_2}\right)-\lambda_3\tau_2r_{01}\left(\frac{E_2}{\tau_2}\right) \notag \\
&~~{\rm s.t.}~~0\leq E_2\leq \tau_2P_{u,\max},~0\leq \tau_2\leq T.\label{eq.sub_part_prob2}
\end{align}
\begin{align}
 & \min_{E_3, \tau_3} ~ E_3+\mu_1\tau_3-\lambda_2\tau_3r_{1}\left(\frac{E_3}{\tau_3}\right)  \notag \\
   & ~~{\rm s.t.}~~0\leq E_3\leq \tau_3P_{h,\max},~0\leq \tau_3\leq T.\label{eq.sub_prob3}
   \end{align}
   \begin{align}
& \min_{l_u\ge 0}  ~~\frac{\kappa_uc_u^3l_u^3}{T^2}-\mu_2l_u \notag \\
&~~~{\rm s.t.}~~ c_ul_u\leq T f_{u,\max}.\label{eq.sub_prob4}
\end{align}
\begin{align}
&\min_{0\leq l_a\leq L}  ~\left(\lambda_2+\lambda_3+\mu_1c_a/f_{a,\max}-\mu_2\right)l_a . \label{eq.sub_prob5}
\end{align}
The optimal solutions to problems \eqref{eq.sub_part_prob1}--\eqref{eq.sub_prob5} are presented in the following Lemmas \ref{lem1}--\ref{lem5}, respectively. As these lemmas can be similarly proved via the Karush-Kuhn-Tucker (KKT) conditions \cite{Boyd2004}, we only show the proof of Lemma \ref{lem1} in Appendix~\ref{AppendixB} and omit the proofs of Lemmas \ref{lem2}--\ref{lem5} for brevity.

\begin{lemma}\label{lem1}
Under given $( {\bm \lambda},{\bm \tau})\in{\cal X}$, the optimal solution $(E_1^*,\tau_1^*,l_h^*)$ to problem \eqref{eq.sub_part_prob1} satisfies
\begin{align}
&E_1^*=P_1^*\tau_1^*,\label{eqn:E1^*}\\
&l_h^*=M_1^*(T-\tau_1^*),\label{eqn:lh^*}\\
\label{eqn:tau1}
&\tau_1^*
\begin{cases}
=T,&{\rm if} ~ \rho_1<0\\
\in[0,T],&{\rm if} ~\rho_1=0\\
=0,&{\rm if} ~ \rho_1> 0,
\end{cases}
\end{align}
where $P_1^*=\left[\frac{\lambda_1B}{\ln2}-\frac{\sigma_1^2}{h_{01}}\right]^{P_{u,\max}}_0$ with $[x]^a_b\triangleq\min\{a,\max\{x,b\}\}$, and
\begin{align}\label{eq.offload_rate_to_helper}
M_1^* \triangleq
\begin{cases}
\left[\sqrt{\frac{\mu_2-\lambda_1}{3\kappa_hc_h^3}}\right]^{\frac{f_{h,\max}}{c_h}}_0 ,&{\rm if} ~ \mu_2-\lambda_1\geq 0\\
0,&{\rm if} ~ \mu_2-\lambda_1<0,
\end{cases}
\end{align}
\begin{align}\label{eqn:rho1}
\rho_1=&\mu_1-\lambda_1r_{01}(P_1^*)+2\kappa_h(c_hM_1^*)^3+\frac{\lambda_1BP_{1}^*{h_{01}}/{\sigma_1^2}}{(1+P_{1}^*{h_{01}}/{\sigma_1^2})\ln2} \nonumber\\
&-\alpha_1 P_{u,\max}+\frac{\beta_1f_{h,\max}}{c_h},
\end{align}
\begin{align}\label{eqn:part:sub1:a2}
\alpha_1\triangleq &
\begin{cases}
0, &{\rm if}~P_1^*<P_{u,\max}\\
\frac{\lambda_1B{h_{01}}/{\sigma_1^2}}{\ln2\left(1+P_1^*h_{01}/{\sigma_1^2}\right)}-1,&{\rm if}~P_1^*=P_{u,\max},
\end{cases}
\end{align}
\begin{align}\label{eqn:part:sub1:b2}
\beta_1\triangleq &
\begin{cases}
0, &{\rm if}~M_1^*<\frac{f_{h,\max}}{c_h}\\
\mu_2-\lambda_1-3\kappa_hc_h^3(M_1^*)^2,&{\rm if}~M_1^*=\frac{f_{h,\max}}{c_h}.
\end{cases}
\end{align}
\end{lemma}
\begin{IEEEproof}
See Appendix~\ref{AppendixB}.
\end{IEEEproof}

\begin{lemma}\label{lem2}
Under given $( {\bm \lambda},{\bm \tau})\in{\cal X}$, the optimal solution $(E_2^*,\tau_2^*)$ to problem (\ref{eq.sub_part_prob2}) satisfies
\begin{align}
&E_2^* = P_2^*\tau_2^*,\label{eqn:E2}\\
\label{eqn:tau2}
&\tau_2^*
\begin{cases}
=T,&{\rm if} ~\rho_2<0\\
\in[0,T],&{\rm if}~\rho_2=0\\
=0,&{\rm if} ~\rho_2> 0,
\end{cases}
\end{align}
where $P_2^*=\left[\frac{\sqrt{v^2-4uw}-v}{2u}\right]^{P_{u,\max}}_0$ with $u=\frac{\ln2}{B}\frac{h_{0}}{\sigma_0^2}\frac{h_{01}}{\sigma_1^2}$,
$v=\frac{\ln2}{B}(\frac{h_{0}}{\sigma_0^2}+\frac{h_{01}}{\sigma_1^2})-(\lambda_2+\lambda_3)\frac{h_{0}}{\sigma_0^2}\frac{h_{01}}{\sigma_1^2}$,
$w=\frac{\ln2}{B}-\lambda_2\frac{h_0}{\sigma_0^2}-\lambda_3\frac{h_{01}}{\sigma_1^2}$, $
\rho_2=\mu_1-\lambda_2r_{0}(P_2^*)+\frac{\lambda_2BP_2^*\frac{h_0}{\sigma_0^2}}{(1+P_2^*\frac{h_0}{\sigma_0^2})\ln2}-\lambda_3r_{01}(P_2^*) +\frac{\lambda_3BP_2^*\frac{h_{01}}{\sigma_1^2}}{(1+P_2^*\frac{h_{01}}{\sigma_1^2})\ln2}-\alpha_2P_{u,\max}$, and
\begin{align*}
\alpha_2=&
\begin{cases}
0, &{\rm if}~P_2^*<P_{u,\max}\\
\frac{\lambda_3B\frac{h_{01}}{\sigma_1^2}}{(1+P_2^*\frac{h_{01}}{\sigma_1^2})\ln2}+\frac{\lambda_2B\frac{h_0}{\sigma_0^2}}{(1+P_2^*\frac{h_0}{\sigma_0^2})\ln2}-1,&{\rm if}~P_2^*=P_{u,\max}.
\end{cases}
\end{align*}
\end{lemma}

\begin{lemma}\label{lem3}
Under given $( {\bm \lambda},{\bm \tau})\in{\cal X}$, the optimal solution $(E_3^*,\tau_3^*)$ to problem (\ref{eq.sub_prob3}) satisfies
\begin{align}
&E_3^* = P_3^*\tau_3^*,\label{eqn:E3}\\
\label{eqn:tau3}
&\tau_3^*
\begin{cases}
=T,&{\rm if} ~\rho_3<0\\
\in[0,T],&{\rm if}~\rho_3=0\\
=0,&{\rm if} ~\rho_3> 0,
\end{cases}
\end{align}
where $P_3^*=\left[\frac{\lambda_2B}{\ln2}-\frac{\sigma_1^2}{h_{1}}\right]^{P_{h,\max}}_0$ and $\rho_3=\mu_1+\frac{\lambda_2BP_3^*\frac{h_{1}}{\sigma_1^2}}{(1+P_3^*\frac{h_{1}}{\sigma_1^2})\ln2}-\lambda_2r_{1}(P_3^*)-\alpha_3P_{h,\max}$ with
\begin{align*}
\alpha_3=&
\begin{cases}
0, &{\rm if}~P_3^*<P_{h,\max}\\
\frac{\lambda_2B{h_{1}}/{\sigma_0^2}}{(1+P_3^*{h_{1}}/{\sigma_0^2})\ln2}-1,&{\rm if}~P_3^*=P_{h,\max}.
\end{cases}
\end{align*}
\end{lemma}

\begin{lemma}\label{lem4}
For given $( {\bm \lambda},{\bm \tau})\in{\cal X}$, the optimal solution $l_u^*$ to problem (\ref{eq.sub_prob4}) is
\begin{align}\label{eqn:l0}
l_u^*=\left[T\sqrt{\frac{\mu_2}{3\kappa_uc_u^3}}\right]^{\frac{Tf_{u,\max}}{c_u}}_0.
\end{align}
\end{lemma}

\begin{lemma}\label{lem5}
For given $( {\bm \lambda},{\bm \tau})\in{\cal X}$, the optimal solution $l_a^*$ to problem (\ref{eq.sub_prob5}) is
\begin{align}
l_a^*\left\{
\begin{array}{ll}
=0,&{\rm if}~\lambda_2+\lambda_3+\mu_1c_a/f_{a,\max}-\mu_2>0\\
\in [0,L],&{\rm if}~\lambda_2+\lambda_3+\mu_1c_a/f_{a,\max}-\mu_2 = 0\\
=L,&{\rm if}~ \lambda_2+\lambda_3+\mu_1c_a/f_{a,\max}-\mu_2<0.
\end{array}\right.\label{eqn:l_a:star}
\end{align}
\end{lemma}

\begin{remark}
Note that in \eqref{eqn:tau1}, \eqref{eqn:tau2}, \eqref{eqn:tau3}, and \eqref{eqn:l_a:star}, if $\rho_i=0$ (for any $i\in\{1,2,3\}$) or $\lambda_2+\lambda_3+\mu_1c_a/f_{a,\max}-\mu_2=0$, then the optimal solution $\tau_i^*$ or $l_a^*$ is non-unique in general. In this case, we choose $\tau_i^* = 0$ and $l_a^*=0$ for the purpose of evaluating the dual function $g( {\bm \lambda}, {\bm \mu})$. Such choices may not be feasible or optimal for problem (P1.1). To tackle this issue, we use an additional step in Section \ref{sec:find:primary} later to find the primal optimal $\tau_i^{\rm opt1}$'s and $l_a^{\rm opt1}$ for problem (P1.1).
\end{remark}

By combining Lemmas~\ref{lem1}--\ref{lem5}, the dual function $g({\bm \lambda},{\bm \mu})$ is evaluated for any given $({\bm \lambda},{\bm \mu})\in{\cal X}$.

\subsection{Obtaining $(\mv \lambda^{\rm opt1}, \mv \mu^{\rm opt1})$ to Maximize $g(\mv \lambda, \mv \mu)$}

Next, we search over $({\bm \lambda},{\bm \mu})\in{\cal X}$ to maximize $g(\mv \lambda, \mv \mu)$ for solving problem $({\rm D1.1})$. Since the dual function $g({\mv \lambda},{\mv \mu})$ is concave but non-differentiable in general, one can use subgradient based methods such as the ellipsoid method \cite{ellipsoid}, to obtain the optimal $\mv \lambda^{\rm opt1}$ and $\mv \mu^{\rm opt1}$ for $({\rm D1.1})$. For the objective function in \eqref{eq.dual_function}, the subgradient with respect to $(\mv \lambda,\mv \mu)$ is
\begin{align*}
&\Big[ l_h^*-\tau_1^*r_{01}\left(\frac{E_1^*}{\tau_1^*}\right), l_a^*-\tau_2^*r_0\left(\frac{E_2^*}{\tau_2^*}\right)-\tau_3^*r_1\left(\frac{E_3^*}{\tau_3^*}\right),\\
&~~ l_a^*-\tau_2^*r_{01}\left(\frac{E_2^*}{\tau_2^*}\right),\sum_{i=1}^3 \tau_i^*+\frac{l_a^*c_a}{f_{a,\max}}-T,L-l_u^*-l_h^*-l_a^*  \Big].
\end{align*}

For the constraints $\mu_1\geq 0$ and $\lambda_i\geq 0$, the subgradients are $\mv e_4$ and $\mv e_i$, $i\in\{1,2,3\}$, respectively, where $\mv e_i$ is the unit vector with one in the $i$-th entry and zeros elsewhere in $\mathbb{R}^{5}$.

\subsection{Optimal Primal Solution to $({\rm P1})$}\label{sec:find:primary}
With $\mv \lambda^{\rm opt1}$ and $\mv \mu^{\rm opt1}$ obtained, it remains to determine the optimal solution to problem $({\rm P1.1})$ (and thus $({\rm P1})$). By replacing $\mv \lambda$ and $\mv \mu$ in Lemmas \ref{lem1}--\ref{lem5} as $\mv \lambda^{\rm opt1}$ and $\mv \mu^{\rm opt1}$, we denote the corresponding $P_i^*$'s, $l_u^*$, and $M_1^*$ as $P_i^{\rm opt1}$'s, $l_u^{\rm opt1}$, and $M_1^{\rm opt1}$, respectively. Accordingly, $\mv P^{\rm opt1} = [P_1^{\rm opt1},P_2^{\rm opt1},P_3^{\rm opt1}]$ corresponds to the optimal solution of $\mv P$ to problem $({\rm P1})$, and $l_u^{\rm opt1}$ corresponds to the optimal solution of $l_u$ to both problems $({\rm P1})$ and $({\rm P1.1})$. Nevertheless, due to the nonuniqueness of $\tau_i^*$'s and $l_a^*$, we implement an additional step to construct the optimal solution of other variables to problem $({\rm P1})$. With ${\mv P}^{\rm opt1}$, $M_1^{\rm opt1}$, and $l_u^{\rm opt1}$, the optimal solution must satisfy $l_h =M_1^{\rm opt1}(T-\tau_1)$ and $E_i =P_i^{\rm opt1} \tau_i$, $i\in\{1,2,3\}$. By substituting them in $({\rm P1})$ or $({\rm P1.1})$, we have the following LP to obtain $\mv \tau^{\rm opt1}$ and $l_a^{\rm opt1}$:
\begin{align}\label{eq.part_prob.2}
\min_{\mv \tau, l_a\ge 0} ~& {\kappa_h(c_hM_1^{\rm opt1})^3(T-\tau_1)}+\sum_{i=1}^3 \tau_iP_i^{\rm opt1} \\ 
{\rm s.t.} ~~~&~M_1^{\rm opt1}(T-\tau_1)\leq \tau_1r_{01}(P_1^{\rm opt1})\notag\\
~&~ l_a\leq \tau_2r_0(P_2^{\rm opt1})+\tau_3r_1(P_3^{\rm opt1})\notag \\
~&~ l_a \leq\tau_2r_{01}(P_2^{\rm opt1})\notag \\
~&~M_1^{\rm opt1}(T-\tau_1) + l_a+ l_u^{\rm opt1}= L\notag \\
~&~ 0\le \tau_i\le T,~\forall i\in\{1,2,3\},~{\rm and}~ \eqref{eqn:T1}.\notag
\end{align}
The LP in (\ref{eq.part_prob.2}) can be efficiently solved by the interior-point method \cite{Boyd2004}. By combining $\mv \tau^{\rm opt1}$, $l_h^{\rm opt1}$, and $l_a^{\rm opt1}$, together with ${\mv P}^{\rm opt1}$ and $l_u^{\rm opt1}$, the optimal solution to problem $({\rm P1})$ is finally found. In summary, we present Algorithm~1 for optimally solving problem $({\rm P1})$ under the partial offloading case in Table I.

\begin{table}[htp]\label{algorithm}
\begin{center}
\caption{Algorithm 1 for Optimally Solving Problem (P1).}  \vspace{0.1cm}
\hrule 
\begin{itemize}
\item[a)] Initialization: Given an ellipsoid ${\cal E}((\mv \lambda, \mv \mu),{\mv A})$ containing $(\mv \lambda^{\rm opt1}, \mv \mu^{\rm opt1})$, where $(\mv \lambda, \mv \mu)$ is the center point of ${\cal E}$ and the positive definite matrix ${\mv A}$ characterizes the size of ${\cal E}$.
\item[b)] {\bf Repeat:}
    \begin{itemize}
    \item[1)]  Obtain ${\mv P}^*,{\mv E}^*, {\mv \tau}^*$, and ${\mv l}^*$ with $(\mv \lambda, \mv \mu)\in{\cal X}$ according to Lemmas~\ref{lem1}--\ref{lem5};
    \item[2)]  Compute the subgradients of $g(\mv \lambda, \mv \mu)$, then update $\mv \lambda$ and $\mv \mu$ using the ellipsoid method\cite{ellipsoid}.
    \end{itemize}
\item[c)] {\bf Until} $\mv \lambda$ and $\mv \mu$ converge with a prescribed accuracy.
\item[d)] {\bf Set} $(\mv \lambda^{\rm opt1}, \mv \mu^{\rm opt1})\gets (\mv \lambda, \mv \mu)$.
\item[e)] {\bf Output}: Obtain $\mv P^{\rm opt1}$ and $l_u^{\rm opt1}$ based on Lemmas~\ref{lem1}--\ref{lem4} by replacing $\mv \lambda$ and $\mv \mu$ as $\mv \lambda^{\rm opt1}$ and $\mv \mu^{\rm opt1}$, and then compute ${\mv \tau}^{\rm opt1}$, $l_h^{\rm opt1}$, and $l_a^{\rm opt1}$ by solving the LP in (\ref{eq.part_prob.2}).
\end{itemize}
\hrule \vspace{0.1cm}\label{algorithm:new0}
\end{center}
\end{table}

\begin{remark}
Based on the optimal solution to (P1) in a semi-closed form, we have the following insights on the optimal joint computation and communication cooperation.
\begin{itemize}
  \item As for local computing, it follows from Lemma \ref{lem4} that the number $l_u^{\rm opt1}$ of task input-bits for local computing generally increases as the block duration $T$ becomes large. This shows that the user prefers locally computing more tasks when the computation latency becomes loose, as will be validated in numerical results later.
  \item As for cooperative computation (i.e., offloading tasks to helper), it is evident that, based on Lemma~\ref{lem1}, the offloading power $P_1^{\rm opt1}$ in the first slot increases as the corresponding channel power gain $h_{01}$ becomes stronger. This is expected, since the marginal energy consumption for offloading from the user to the helper reduces in this case, and thus the user prefers offloading more tasks to the helper for cooperative computation.
  \item As for cooperative communication (i.e., offloading to AP), it is observed from Lemma~\ref{lem2} and \ref{lem3} that the offloading power $P_2^{\rm opt1}$ in the second slot is dependent on both $h_{01}$ and $h_0$, while $P_3^{\rm opt1}$ in the third slot increases as $h_1$ becomes large.
\end{itemize}
\end{remark}

\section{Optimal Solution to ({\rm P2})}\label{sec:binary}

In this section, we develop an efficient algorithm to optimally solve problem (P2) in the binary offloading case.

Due to the constraints in \eqref{eqn:L} and \eqref{eq.part_prob2.b}, there exist in total three computation modes for the user's task execution, i.e., the local computing mode (with $l_u=L$ and $l_h=l_a=0$), the computation cooperation mode (with $l_h=L$ and $l_u=l_a=0$), and the communication cooperation mode (with $l_a=L$ and $l_u=l_h=0$). In the following, we first obtain the energy consumption under each of the three computation modes, and then choose the best mode with the minimum energy consumption as the optimal solution to problem (P2).

\subsection{Computation Modes for Binary Offloading Case}
\subsubsection{Local Computing Mode}\label{modeu}
The local computing mode is feasible only when $l^{(2)}_{u,\max}\geq L$, with $l^{(2)}_{u,\max}$ given in \eqref{eq.bin_lumax}. In this case, by substituting $l_u=L$ and $l_h=l_a=0$ in (P2), we have the optimal transmit power and time allocation as ${\bm P}=\mv 0$ and $\bm \tau=\mv 0$,  respectively. Therefore, the minimum energy consumption by the user in this mode is
\begin{align}\label{eq:binary:user:E}
E_u^{\rm opt2}=\frac{\kappa_uc_u^3L^3}{T^2}.
\end{align}

\subsubsection{Computation Cooperation Mode}\label{modeh}
The computation cooperation mode is feasible only when $l^{(2)}_{h,\max}\geq L$, with $l^{(2)}_{h,\max}$ given in \eqref{eq.bin_lhmax}. Substituting $l_h=L$ and $l_u=l_a=0$ into (P2), it then follows that $P_2=P_3=0$ and $\tau_2=\tau_3=0$. Consequently, problem (P2) is reduced as
\begin{subequations}\label{eq:pro:binary:HE}
\begin{align}
\min_{P_1, \tau_1} &~\tau_1P_1+\frac {\kappa_hc_h^3L^3}{(T-\tau_1)^2} \label{eq:pro:binary:HE.a}\\
{\rm s.t.} &~ L \leq \tau_1r_{01}(P_1) \label{eq:pro:binary:HE.b}\\
&~ {c_h}L\le{(T-\tau_1)f_{h,\max}} \label{eq:pro:binary:HE.c}\\
&~0\leq \tau_1\leq T,~0\leq P_1\leq P_{u,\max}. \label{eq:pro:binary:HE.d}
\end{align}
\end{subequations}
Note that at the optimality of \eqref{eq:pro:binary:HE}, the constraint \eqref{eq:pro:binary:HE.b} must be tight. It thus follows that $P_1 =\left(2^{\frac{L}{B\tau_1}}-1\right)\frac{\sigma_1^2}{h_{01}}$. Accordingly, problem \eqref{eq:pro:binary:HE} is further reduced as the following univariable convex optimization problem:
\begin{align} \label{eq:pro:binary:HE^*}
\tau_1^{\rm opt2} \triangleq \arg \min_{\tau_1} &~\left(2^{\frac{L}{B\tau_1}}-1\right)\frac{\tau_1\sigma_1^2}{h_{01}}+\frac {\kappa_hc_h^3L^3}{(T-\tau_1)^2}\\
{\rm s.t.} & ~\left(2^{\frac{L}{B\tau_1}}-1\right)\frac{\sigma_1^2}{h_{01}} \le P_{u,\max}\notag\\
&~~0\leq \tau_1\leq T-\frac{c_hL}{f_{h,\max}},\notag
\end{align}
where the optimal solution $\tau_1^{\rm opt2}$ to problem \eqref{eq:pro:binary:HE^*} can be efficiently found via a bisectional search procedure \cite{Boyd2004}. With $\tau_1^{\rm opt2}$ obtained, the sum energy consumption at the user and the helper is given by
\begin{align}\label{eq:pro:binary:HE^**}
E_h^{\rm opt2} = \left(2^{\frac{L}{B\tau_1^{\rm opt2}}}-1\right)\frac{\tau_1^{\rm opt2}\sigma_1^2}{h_{01}}+\frac {\kappa_hc_h^3L^3}{(T-\tau_1^{\rm opt2})^2}.
\end{align}

\subsubsection{Communication Cooperation Mode}\label{modeap}
The communication cooperation mode is feasible only when $l_{a,\max}^{(2)} \ge L$ with $l_{a,\max}^{(2)}$ given in \eqref{eq.bin_lamax}. With $l_a=L$ and $l_u=l_h=0$, it follows that $P_1=0$ and $\tau_1=0$. Therefore, problem (P2) is re-expressed as
\begin{align}\label{eq:pro:binary:AP}
\min_{\tau_{2},\tau_3, P_2,P_3}&~~\tau_2P_2+\tau_3P_3\\
{\rm s.t.}~~&~~L\leq \min\left(\tau_2r_0(P_2)+\tau_3r_1(P_3),~\tau_2r_{01}(P_2)\right) \notag\\
&~~ \tau_2+\tau_3+Lc_a/f_{a,\max} \leq T \notag\\
&~~ 0\leq\tau_i\leq T, ~\forall i\in\{2,3\}  \notag\\
&~~ 0\leq P_2\leq P_{u,\max},~0\leq P_3\leq P_{h,\max}.\notag
\end{align}
Similarly as for problem (P1), problem \eqref{eq:pro:binary:AP} can be optimally solved by Algorithm 1 by setting $l_u = 0$, $l_h = 0$, $l_a = L$, and $\tau_1 = 0$. We denote $(\tau_2^{\rm opt2}, \tau_2^{\rm opt2}, P_1^{\rm opt2}, P_2^{\rm opt2})$ as the optimal solution to problem \eqref{eq:pro:binary:AP}. Therefore, we obtain the energy consumption for this mode as
\begin{align}\label{eq:pro:binary:AP**}
E_a^{\rm opt2} = \tau_2^{\rm opt2} P_2^{\rm opt2} + \tau_3^{\rm opt2} P_3^{\rm opt2}.
\end{align}

\subsection{Computation Mode Selection}

By comparing $E_u^{\rm opt2}$, $E_h^{\rm opt2}$, and $E_a^{\rm opt2}$, we determine the optimal computation mode for problem (P2) as the one with the minimum energy consumption. Accordingly, the optimal $l_a$, $l_u$, and $l_h$ are decided, and the corresponding computation and communication resources allocation for the selected computation mode becomes the optimal solution to problem (P2). As a result, problem (P2) in the binary offloading case is optimally solved.

\section{Numerical Results}\label{sec:simulation}

In this section, numerical results are provided to evaluate the performance of the proposed joint computation and communication cooperation design for both partial and binary offloading cases, as compared to the following five benchmark schemes without such a joint design.

\begin{itemize}
  \item {\it Local computing}: The user executes the computation tasks locally by itself. The energy consumption for local computing is obtained as $E_u^{\rm opt2}$ in \eqref{eq:binary:user:E}.
  \item {\it Computation cooperation with partial offloading \cite{xuchenfog}}: The computation tasks are partitioned into two parts for the user's local computing and offloading to the helper, respectively. This corresponds to solving problem (P1) by setting $l_a=0$ and $\tau_2=\tau_3=0$.
  \item {\it Communication cooperation with partial offloading \cite{Huang17}}: The computation tasks are partitioned into two parts for the user's local computing and offloading to the AP, respectively. The offloading is assisted by the helper's communication cooperation as a DF relay. This corresponds to solving problem (P1) by setting $l_h=0$ and $\tau_2+\tau_3=T$.
  \item {\it Computation cooperation with binary offloading}: The user offloads all the computation tasks to the helper for remote execution. The energy consumption corresponds to $E_h^{\rm opt2}$ in \eqref{eq:pro:binary:HE^**}.
  \item {\it Communication cooperation with binary offloading}: The user can only offload its computation tasks to the AP assisted by the helper's communication cooperation. The energy consumption corresponds to $E_a^{\rm opt2}$ in (\ref{eq:pro:binary:AP**}) based on \eqref{eq:pro:binary:AP}.

\end{itemize}

In the simulation, we consider that the user and the AP are located with a distance of $250$ meters (m) and the helper is located on the line between them. Let $D$ denote the distance between the user and the helper. The pathloss between any two nodes is denoted as $\beta_0\left( {d}/{d_0}\right)^{-\zeta}$, where $\beta_0=-60$~dB corresponds to the path loss at the reference distance of $d_0=10$~m, $d$ denotes the distance from the transmitter to the receiver, and $\zeta=3$ is the pathloss exponent.
Furthermore, we set $B=1$~MHz, $\sigma_0^2=\sigma_1^2=-70$~dBm, $c_u=c_h=10^3$ cycles/bit \cite{Wang17}, $\kappa_u=10^{-27}$ \cite{Huang17}, $\kappa_h =0.3 \times 10^{-27}$, $P_{u,\max}=P_{h,\max}=40$~dBm, $f_{u,\max}=2$~GHz, $f_{h,\max}=3$~GHz, and $f_{a,\max}=5$~GHz.

\begin{figure}
\centering
    \includegraphics[width=8.5cm]{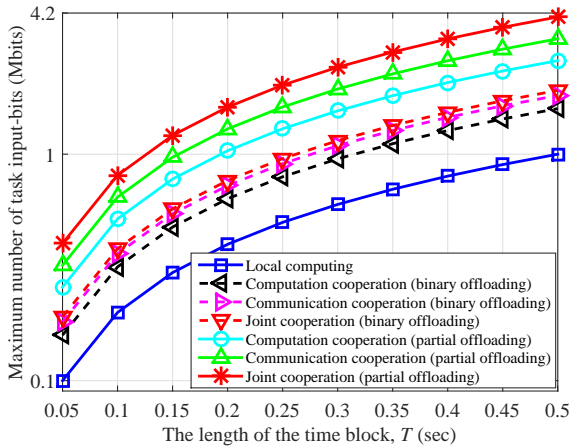}
\caption{The maximal number of task input-bits versus the block length.} \label{fig_sim_fea_lmax}
\end{figure}

Fig.~\ref{fig_sim_fea_lmax} shows the maximum number of task input-bits versus the block duration $T$, where $D=20$~m. It is observed that for both partial and binary offloading cases in this setup, the computation cooperation and communication cooperation schemes achieve higher computation capacity than the local computing benchmark. Furthermore, for the binary or partial offloading, the communication cooperation scheme is observed to outperform the corresponding computation cooperation schemes, due to the higher computation capacities at the MEC server. In addition, our proposed joint cooperation design is observed to achieve the highest computation capacity by exploiting both benefits.

\begin{figure}
\centering
    \includegraphics[width=8.5cm]{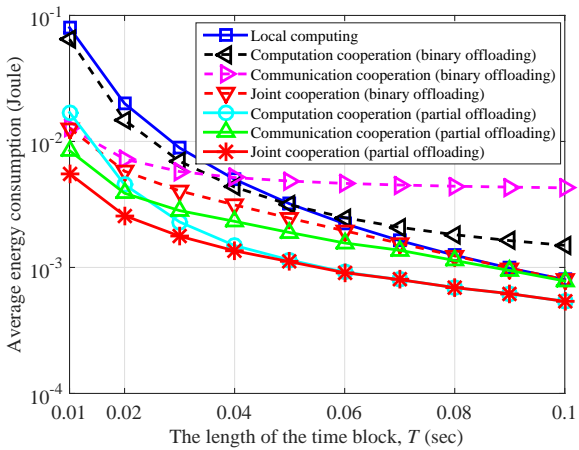}
\caption{The average energy consumption versus the block length.} \label{fig_sim_T1}
\end{figure}

Fig.~\ref{fig_sim_T1} shows the average energy consumption versus the block length $T$, where $L=0.02$~Mbits and $D=120$~m. The proposed joint cooperation scheme is observed to achieve the minimum energy consumption for both the partial and binary offloading cases, respectively. In addition, we have the following observations.
\begin{itemize}
  \item The average energy consumption by all the schemes decreases as $T$ increases. By comparing the partial and binary offloading, the computation and/or communication cooperation approach in the former case achieves more significant energy reduction than the corresponding one in the latter case. This indicates the benefit of task partitions in energy saving for MEC.
  \item In the binary offloading case, when $T$ is small (e.g., $T < 0.035$ sec), the communication cooperation scheme achieved a lower energy consumption than the local computing and the computation cooperation schemes. When $0.035$~sec $< T \le 0.05$~sec, the computation cooperation scheme outperforms the other two schemes. As $T$ becomes large (e.g., $T > 0.05$ sec), the local computing scheme is beneficial.
  \item In the partial offloading case, the communication cooperation scheme achieves a lower energy consumption than the computation cooperation scheme when $T$ is small (e.g., $T<0.025$~sec), while the reverse is true when $T$ becomes large. Also, the computation cooperation and communication cooperation schemes both outperform the local computing one. This is because that the two cooperation schemes additionally exploit computation resources at the helper and the AP, respectively.
\end{itemize}

\begin{figure}
\centering
\includegraphics[width=8.5cm]{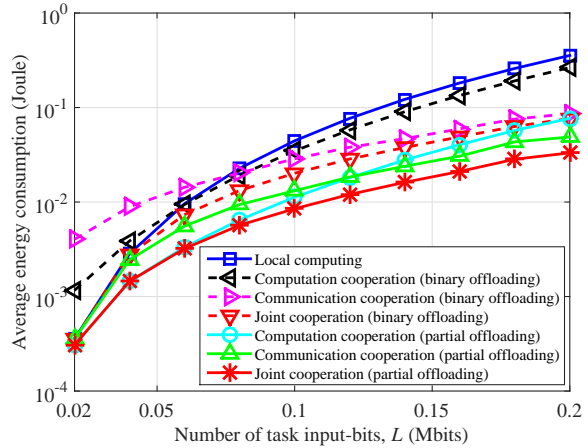}
\caption{The average energy consumption versus the number of task input-bits.} \label{fig_sim_L1}
\end{figure}

Fig.~\ref{fig_sim_L1} shows the average energy consumption versus the number of task input-bits $L$, where $T=0.15$~sec and $D=120$~m. We have generally similar observations in Fig.~\ref{fig_sim_L1} as in Fig.~\ref{fig_sim_T1}. Specifically, it is observed that at small $L$ values (e.g., $L < 0.06$ Mbits), the local computing is observed to achieve a similar performance as the proposed joint cooperation scheme with binary offloading, since in this case, the local computing is sufficient to execute the computation task input-bits. As $L$ increases, the joint computation and communication cooperation is observed to achieve significant performance gain in terms of energy reduction.

\begin{figure}
\centering
    \includegraphics[width=8.5cm]{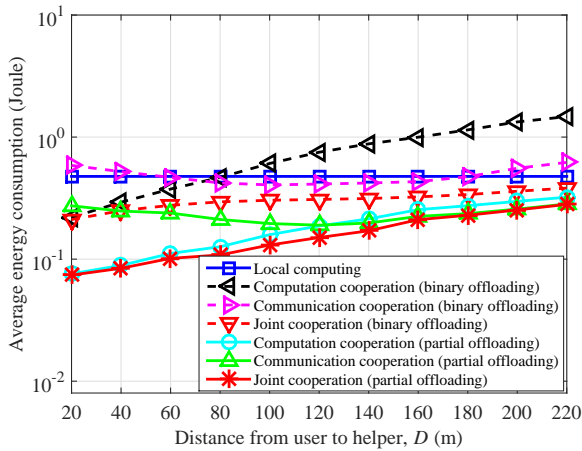}
\caption{The average energy consumption versus the distance between the user and the helper.} \label{fig_sim_D}
\end{figure}

Fig.~\ref{fig_sim_D} shows the average energy consumption versus $D$ in both the partial and binary offloading cases, where $T=0.3$~sec and $L=0.5$~Mbits.
As expected, the average energy consumption by the local computing scheme remains unchanged for all $D$ values. As $D$ becomes larger, the average energy consumption by the communication cooperation scheme is observed to first decrease and then increase, while that by the computation cooperation scheme is observed to increase monotonically. This is because that as $D$ increases, the channel gain between the user and the helper becomes smaller, while that between the helper and the AP becomes stronger; therefore, such a change benefits the offloading from the helper to the AP, but incurs increased offloading energy consumption from the user to the helper. Furthermore, the proposed joint cooperation scheme is observed to achieve significant gains over these benchmark schemes at all $D$ values.

\section{Concluding Remarks}\label{sec:conclusions}

In this paper, we proposed a novel joint computation and communication cooperation approach for improving the MEC performance, where nearby helper nodes share the computation and communication resources to help actively-computing user nodes. By considering a basic three-node model under a four-slot cooperation protocol, we developed an energy-efficient design framework for both partial and binary offloading cases. We minimized the total energy consumption at both the user and the helper subject to the computation latency constraint, by jointly optimizing the task partition, as well as the computation and communication resources allocation. Based on convex optimization techniques, we presented an efficient algorithm to obtain the optimal solution in the partial offloading case. Computation mode selection was then applied for optimally solving the problem in the binary offloading case. Extensive numerical results demonstrated the merit of the proposed joint computation and communication cooperation scheme over alternative benchmarks. It is our hope that this paper sheds new light on how to optimally design multi-resource user cooperation to improve the operation efficiency of MEC. Due to the space limitation, there have been various important issues that have not been addressed in this paper, which are discussed as follows to motivate future work.
\begin{itemize}
\item
Although this paper considered the basic setup with one user and one helper, our results are extendable to the more general case with multiple users and helpers. For example, in this case, we can employ a user-helper pairing procedure to pair each user with one helper, such that the helper can use the proposed joint communication and computation cooperation design to help the computation of the paired user. Furthermore, to fully utilize the computation and communication resources at multiple helpers, each user can offload the tasks to multiple helpers simultaneously for parallel execution, and multiple helpers can also cooperatively relay the user's tasks to the AP (e.g., via the collaborative beamforming). However, how to efficiently pair multiple users and helpers, and efficiently design the multiuser computation offloading and collaborative relaying are new and generally difficult problems worthy of further investigation.
\item
This paper assumed that the user and helper have the common interest in improving the MEC system energy efficiency, such that the centralized resource allocation is employed for performance optimization. In practice, however, the helper can have self-interest. In this case, how to design incentive mechanisms (such as monetary and credit-based ones) and distributed algorithms to motivate the helper to participate in the joint cooperation is an interesting problem worth pursuing in the future.
\end{itemize}

\appendix

\subsection{Proof of Lemma~\ref{lem0}}\label{AppendixA}
Note that for ${{\cal L}}({\bm E},{\bm \tau}, {\bm l}, {\bm \lambda}, {\bm \mu})$ in~(\ref{eq.dual_function}), there exist two terms $(\lambda_1-\mu_2)l_h$ and $(\lambda_2+\lambda_3+\mu_1c_a/f_{a,\max}-\mu_2)l_a$.
Suppose $\lambda_1-\mu_2<0$ (or $\lambda_2+\lambda_3+\mu_1c_a/f_{a,\max}-\mu_2<0$). Then ${\cal L}({\mv E},{\mv \tau},{\mv l},{\mv \lambda},{\mv \mu})$ becomes negative infinity when $l_h\to +\infty$ (or $l_a\to +\infty$). This implies that the dual function $g(\mv \lambda,\mv \mu)$ is unbounded from below in this case. Hence, it requires that $\lambda_1-\mu_2\geq 0$ and $\lambda_2+\lambda_3+\mu_1c_a/f_{a,\max}-\mu_2 \geq 0$ to guarantee $g(\mv \lambda,\mv \mu)$ to be bounded from below.  Lemma~\ref{lem0} thus follows.

\subsection{Proof of Lemma~\ref{lem1}}\label{AppendixB}

As problem (\ref{eq.sub_part_prob1}) is convex and satisfies the Slater's condition, strong duality holds between problem (\ref{eq.sub_part_prob1}) and its dual problem. Therefore, one can solve this problem by applying the KKT conditions~\cite{Boyd2004}. The Lagrangian of problem (\ref{eq.sub_part_prob1}) is given by
\begin{align*}
{\cal L}_1=&E_1+\mu_1\tau_1-\lambda_1\tau_1r_{01}(\frac{E_1}{\tau_1})-\mu_2l_h+\lambda_1l_h+\frac{\kappa_hc_h^3l_h^3}{(T-\tau_1)^2}\\
&-a_1E_1+\alpha_1(E_1-\tau_1 P_{u,\max})-b_1\tau_1+b_2(\tau_1-T)\\
&-d_1l_h+\beta_1\left(l_h-\frac{(T-\tau_1)f_{h,\max}}{c_h}\right),
\end{align*}
where $a_1$, $\alpha_1$, $b_1$, $b_2$, $d_1$, and $\beta_1$ are the non-negative Lagrange multipliers associated with $E_1\geq0$, $E_1\leq \tau_1 P_{u,\max}$, $\tau_1\geq0$, $\tau_1\leq T$, $l_h\geq0$, and $l_h\leq {(T-\tau_1)f_{h,\max}}/{c_h}$, respectively.

Based on the KKT conditions, it follows that
\begin{subequations}\label{eq.kkt}
\begin{align}
a_1E_1& =0, ~\alpha_1 (E_1-\tau_1 P_{u,\max})=0, ~b_2(\tau_1-T)=0\\
b_1\tau_1&=0,~d_1l_h=0,~\beta_1\left(l_h-\frac{(T-\tau_1)f_{h,\max}}{c_h}\right)=0\\
\frac{\partial {\cal L}_1}{\partial E_1}&=1-\frac{\lambda_1B\frac{h_{01}}{\sigma_1^2}}{\ln2\left(1+\frac{E_1}{\tau_1}\frac{h_{01}}{\sigma_1^2}\right)}-a_1+\alpha_1 =0\\
\frac{\partial {\cal L}_1}{\partial \tau_1}&= \frac{2\kappa_hc_h^3l_h^3}{(T-\tau_1)^3}-\lambda_1B\log_2 \left( 1+\frac {E_1}{\tau_1}\frac{h_{01}}{\sigma_1^2} \right)+\frac{\beta_1 f_{h,\max}}{c_h}\notag \\
&+\mu_1+\frac{\lambda_1B\frac{h_{01}}{\sigma_1^2}\frac{E_1}{\tau_1}}{\ln2\left(1+\frac{E_1}{\tau_1}\frac{h_{01}}{\sigma_1^2}\right)}-b_1+b_2+\alpha_1 P_{u,\max}=0\\
\frac{\partial {\cal L}_1}{\partial l_h}& =\frac{3\kappa_hc_h^3l_h^2}{(T-\tau_1)^2}-\mu_2+\lambda_1-d_1+\beta_1=0,
\end{align}
\end{subequations}
where (\ref{eq.kkt}a) and (\ref{eq.kkt}b) denote the complementary slackness condition, (\ref{eq.kkt}c), (\ref{eq.kkt}d) and (\ref{eq.kkt}e) are the first-order derivative conditions of ${\cal L}_1$ with respect to $E_1$, $\tau_1$, and $l_h$, respectively.
Therefore, we have (\ref{eqn:E1^*}) based on (\ref{eq.kkt}c), and (\ref{eqn:lh^*}) holds due to (\ref{eq.kkt}e). Furthermore, based on (\ref{eq.kkt}c), (\ref{eq.kkt}d), and (\ref{eq.kkt}e) and with some manipulations, we have \eqref{eqn:part:sub1:a2} and \eqref{eqn:part:sub1:b2}.

Furthermore, by substituting (\ref{eqn:E1^*}) and (\ref{eqn:lh^*}) into (\ref{eq.kkt}d) and assuming $\rho_1=b_2-b_1$, we have $\rho_1$ in \eqref{eqn:rho1}.
Hence, the optimal $\tau_1^*$ is given in (\ref{eqn:tau1}). This lemma is thus proved.

\end{document}